\documentclass[amsmath,amssymb,superscriptaddress,reprint]{revtex4-1}
\usepackage{graphicx}
\usepackage[dvipdfmx]{color}
\usepackage{dcolumn}
\usepackage{bm}
\usepackage{txfonts}
\usepackage{multirow}
\usepackage[utf8]{inputenc}
\usepackage[T1]{fontenc}
\usepackage{mathptmx}
\usepackage{etoolbox}
\usepackage[hypertex,colorlinks=true,linkcolor=black,citecolor=blue,filecolor=blue,urlcolor=blue,setpagesize=false,nesting=true]{hyperref}

\makeatletter
\def\@email#1#2{%
 \endgroup
 \patchcmd{\titleblock@produce}
  {\frontmatter@RRAPformat}
  {\frontmatter@RRAPformat{\produce@RRAP{*#1\href{mailto:#2}{#2}}}\frontmatter@RRAPformat}
  {}{}
}%
\makeatother

\begin{document}
\title{Extended Dynamical Kubo-Toyabe Relaxation for $\mu$SR study of Ion Dynamics: \\An Introduction}
\author{Takashi U. Ito}\thanks{ito.takashi15@jaea.go.jp}
\affiliation{Advanced Science Research Center, Japan Atomic Energy Agency, Tokai-mura, Naka-gun, Ibaraki 319-1195, Japan}
\author{Ryosuke Kadono}\thanks{ryosuke.kadono@kek.jp}
\affiliation{Muon Science Laboratory, Institute of Materials Structure Science, High Energy Accelerator Research Organization (KEK), Tsukuba, Ibaraki 305-0801, Japan}

\begin{abstract}%
In the analysis of the ion diffusion in metal oxides based on muon spin rotation and relaxation ($\mu$SR), the dynamical Kubo-Toyabe (dKT) function has been routinely used to deduce the jump frequency of ions.  This is based on the two beliefs: (1) the fluctuations of the internal magnetic field ${\bm H}(t)$ are determined solely by the relative motion of the muons to the surrounding ions, and (2) the muons are immobile due to bonding to the oxygen. However, these are not necessarily trivial, and we addressed their credibility by developing an extended dKT function corresponding to the realistic situation that only a part of the ions surrounding muon are involved in a single fluctuation of ${\bm H}(t)$ in the ion diffusion, and investigated its behavior in detail. The results show that the new function exhibits qualitatively different behavior from the dKT function, and that it provides a way to determine whether muons or ions are in motion, as well as a means for quantitative analysis based on the assumption of immobile muons. As a typical example, we examine the earlier $\mu^\pm$SR results on Na$_x$CoO$_2$ and demonstrate that the internal field fluctuations observed in $\mu^+$SR are dominated by muon self-diffusion, in contrast to previous interpretations.
\end{abstract}

\maketitle

\section{Introduction}
Muon spin rotation relaxation ($\mu$SR) is an experimental technique analogous to nuclear magnetic resonance (NMR) in time domain, and is widely used as a microscopic probe for materials science research \cite{MSRPSI,MSR}.  As elementary particles, muons exist in either positively or negatively charged states ($\mu^+$ or $\mu^-$), and $\mu$SR experiment can be performed with either particle (sometimes explicitly called $\mu^+$SR or $\mu^-$SR). However, despite recent efforts to overcome various difficulties associated with $\mu^-$SR,  $\mu^+$ remains a major player for materials science applications. Therefore, unless otherwise specified, $\mu^+$s will be referred to as muons as below.

\begin{figure}[t]
	\centering
  \includegraphics[width=0.75\linewidth]{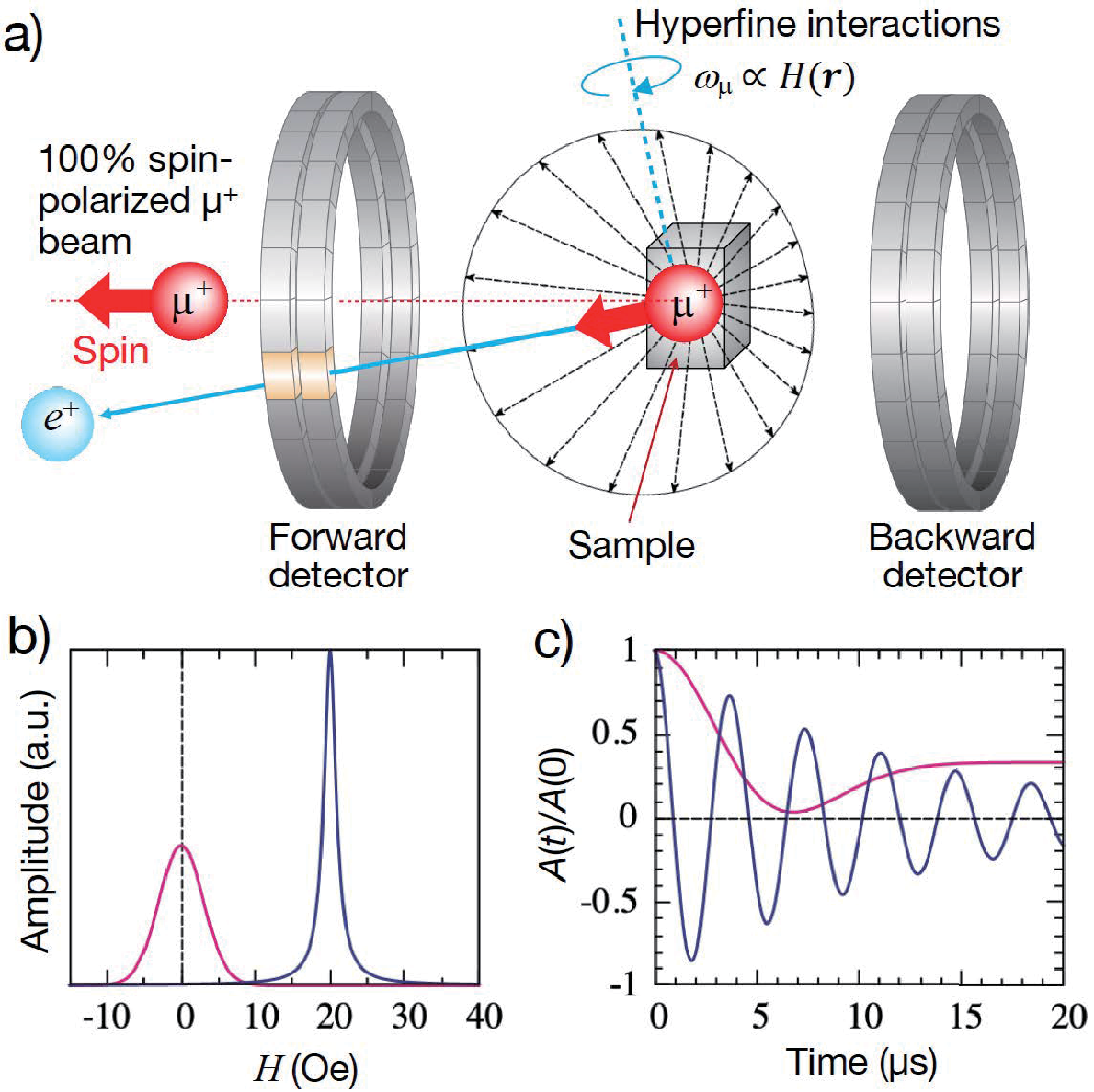}
\caption{(a) Schematics of the $\mu$SR experiment: spin-polarized muons are implanted into a sample, and the time evolution of muon spin polarization is monitored by measuring time-dependent decay-positron asymmetry $A(t)$, taking advantage of the preferential emission of positrons in the muon spin direction. The spin rotation (Larmor precession) frequency $\omega_\mu$ is proportional to the local magnetic field ${\bm H}({\bm r})$ at the muon site(s) ${\bm r}$ (hyperfine interactions in the general sense). (b) Examples of magnetic field distribution felt by diamagnet Mu: red $=\langle{\bm H}({\bm r})\rangle=0$, blue $=\langle{\bm H}({\bm r})\rangle=20$ Oe, and (c) the corresponding time spectrum [normalized by $A(0)$].}
\label{fig1}
\end{figure}

As illustrated in Fig.~\ref{fig1}(a), The analysis of material properties by $\mu$SR begins with the implantation of spin-polarized muons into a sample of interest. Due to its positive charge and mass close to 1/9th of proton, the muon behaves as a light radioisotope of proton in matter: it stops at specific positions (interstitial sites) in the crystalline lattice, then probes the local magnetic field (while undergoing jump motion in some cases), and finally decays into an energetic positron and two neutrinos with an average lifetime of $\tau_\mu\simeq2.197$ $\mu$s. Since the positron is preferentially emitted in the muon spin direction upon its decay, the spatial asymmetry ($A(t)$) in the positron distribution can be used to monitor the time evolution of spin polarization (called $\mu$SR ``time spectra'') for a period up to 10$\tau_\mu$ ($\sim$20 $\mu$s). While the information obtained in this way is similar to that by NMR which uses polarized nuclear spins as probes, there are considerable differences in sensitivity to local magnetic fields and in the time range of observation between them. In particular, since the gyromagnetic ratio of muons ($\gamma_\mu=2\pi\times 13.553$ kHz/Oe) is about 3.18 times larger than that of protons, $\mu$SR is more sensitive to local magnetic fields originating from nuclear magnetic moments around muons.

As one of the applications that take such advantage of $\mu$SR, ion diffusion in battery materials is a subject of active research in the recent years \cite{Sugiyama:09,Mansson:13,Baker:11,Sugiyama:20,Ohishi:22a,Ohishi:22b,Umegaki:22,Ohishi:23,Nocerino:24}. The diffusive motion of ions with sizable nuclear magnetic moments (and sufficient natural abundance) is assumed to exert fluctuating internal fields to muons, and the corresponding $\mu$SR spectra are analyzed by the dynamical Kubo-Toyabe (dKT) function that incorporates the effect of fluctuations in the local magnetic field based on the ``strong collision'' approximation. This function was initially discovered in the theoretical investigation of NMR in the hypothetical zero-field (ZF) limit (and without electric quadrupole interaction) \cite{Kubo:66}, and later became widely applied after it was found to properly describe ZF muon spin relaxation observed in metals where muons exhibit self-diffusion \cite{Hayano:79}. 

\begin{figure}[t]
	\centering
  \includegraphics[width=0.95\linewidth]{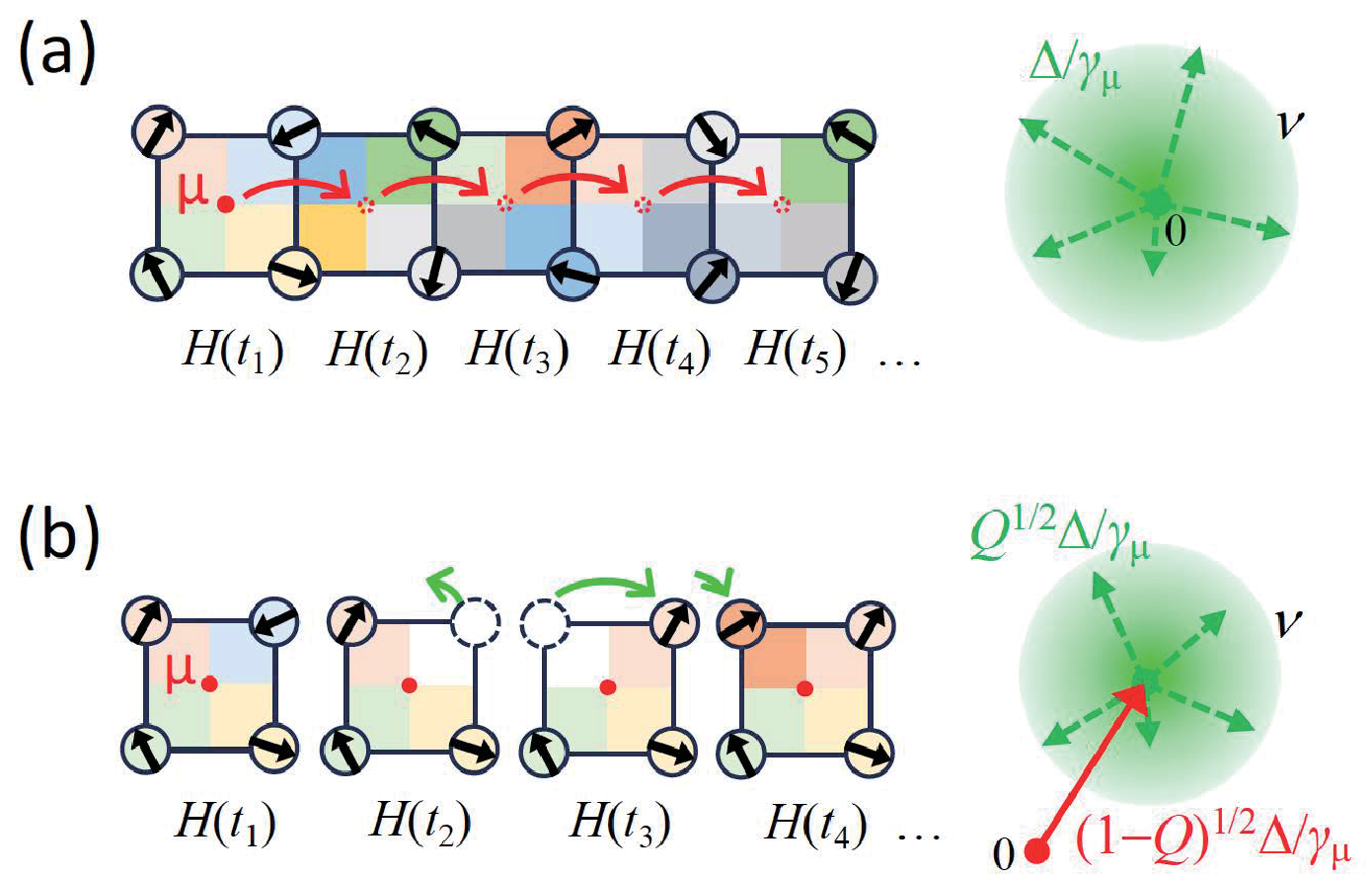}
\caption{Schematics of the time series $t_i$ of muon/ion diffusive motion and the associated fluctuations of the internal magnetic fields ${\bm H}(t_i)$ at muon sites (red spots) \cite{Ito:24}. The black arrows represent nuclear spins of ions, and their respective contributions to ${\bm H}(t_i)$ are represented by colored tiles.  (a) The muon diffuses through the interstitial sites surrounded by stationary ions. (b) The ions diffuse arround the immobile muon. ${\bm H}(t_i)$ in (b) is represented by the vector sum of quasistatic and fluctuating fields indicated by red (solid) and green (dashed) arrows.}
\label{fig2}
\end{figure}

The basic assumptions in deriving the dKT function are that the local magnetic field ${\bm H}({\bm r})$ (which varies randomly at different muon sites) is approximately given by a Gaussian density distribution $n({\bm H})$ with standard deviation $\Delta/\gamma_\mu$ [red line in Fig.~\ref{fig1}(b)], and that ${\bm H}(t)$ felt by individual muons at a given time $t$ is considered to vary randomly with frequency $\nu$ while maintaining $n({\bm H})$ as a whole. This corresponds well to the process of muons jumping with a rate $\nu$  through numerous equivalent sites and probing ${\bm H}({\bm r},t)$ in various orientations and magnitudes over a full range [Fig.~\ref{fig2}(a)]. 

In the strong collision approximation, no explicit distinction is made as to whether the change in ${\bm H}$ is due to muon or ion motion. This is presumably why the effect of fluctuations in ${\bm H}$ due to ion diffusion is also believed to be described by the dKT function, leading to its routine application for analyzing $\mu$SR spectra observed in battery materials to deduce fluctuation rate which is regarded as the jumping frequency of ions \cite{Mansson:13}.

However, a closer examination of the actual ion diffusion situation reveals that the assumptions that make the dKT function applicable are not always fulfilled.  For example, if we consider layered cobaltate compounds $A_x$CoO$_2$ ($A=$ Li, Na) which are typical ion battery cathode materials, muons randomly placed at interstitial sites in these compounds will simultaneously probe local magnetic fields originating from both $^{6/7}$Li/$^{23}$Na  and $^{59}$Co nuclear magnetic moments with Co ions fixed at lattice points. When the Co nuclear magnetic moments are static on the $\mu$SR time scale (as is expected for non-magnetic Co$^{3+}$ ions), then the local magnetic field felt by individual muons is a vector sum of contributions from diffusing Li/Na and stationary Co ions. This means that the autocorrelation of ${\bm H}(t)$ is non-zero when averaged over a long time period, violating the condition for applying the dKT function Fig.~\ref{fig2}(b)]. Even in the case where all the ions around the muon are mobile, it is highly unlikely that a single change in ${\bm H}(t)$ is due to their simultaneous jump, and the situation is expected to be very different from the case of muon self-diffusion. Therefore, it is not clear whether the spin relaxation function expected in such a case would be the same as the dKT function, and a detailed theoretical investigation is needed.

In this paper, we introduce a new theoretical model for muon spin relaxation that incorporates quasistatic contributions from ions fixed at lattice points as well as dynamic contributions from ions diffusing in the solid \cite{Ito:24}. First, the conventional analysis of ion dynamics using the dKT function is reviewed. Then this new model is introduced as an extension of the dKT model, where the two relaxation functions are compared in detail to show the qualitative difference between them. In particular, the strong influence of the quasistatic contribution to the lineshape that allows distinction between ion dynamics and muon self-diffusion will be demonstrated through comparison with the earlier $\mu$SR data in Na$_x$CoO$_2$.

\section{Problems with ${\bf d}$KT Function in Ion Dynamics Analysis}
Since the first report of $\mu$SR studies on a battery cathode material Li$_x$CoO$_2$ \cite{Sugiyama:09}, the data analysis presented therein seems to serve as a standard procedure until now, where a set of $\mu$SR spectra under different longitudinal magnetic fields ($H_{\rm LF}$, including zero field) are obtained at each temperature point, which are then subjected to analysis by the least-square fit using the dKT function $G_z^{\rm KT}(t;\Delta,\nu,H_{\rm LF})$ to deduce the linewidth $\Delta$ and the fluctuation rate $\nu$ as common parameters (see below). While the literary meaning of $\nu$ is the fluctuation frequency of ${\bm H}(t)$ felt by the muon \cite{Hayano:79}, $\nu$ is mostly interpreted as the jump frequency of the ions under the assumption that the muon is immobile regardless of temperature.

Here, let us reiterate the details of the dKT model that are the key to this analysis procedure. The model is based on the general assumption that the distribution of ${\bm H}$ obeys a three-dimensional Gaussian distribution 
\begin{equation}
n(H_\alpha)=\frac{\gamma_\mu}{\sqrt{2\pi}\Delta}e^{-\frac{\gamma_\mu^2H_\alpha^2}{2\Delta^2}},\:\:(\alpha=x,y,z)
\end{equation}
with mean zero and standard deviation $\Delta/\gamma_\mu=\langle H_\alpha({\bm r})^2\rangle^{1/2}$.
 For the static case ($\nu = 0$), by calculating the Larmor precession of the muon spin under ${\bm H}$ (projected to the $z$ axis parallel with the initial muon polarization)  and taking a weighted average over $n({\bm H})$, the static KT function is obtained in an analytic form as below \cite{Hayano:79},
\begin{equation}
g_z^{\rm KT}(t;\Delta)=\frac{1}{3}+\frac{2}{3}(1-\Delta^2t^2)e^{-\frac{1}{2}\Delta^2t^2}.\label{sKT}
\end{equation}
Using this function as a starting point, the dKT function can be obtained by taking into account the fluctuations of ${\bm H}(t)$, for which a Gaussian-Markov process \cite{Comt} was adopted as a stochastic model by Kubo and Toyabe \cite{Kubo:66}. Assuming the correlation time of ${\bm H}(t)$ is $\nu^{-1}$, its autocorrelation function takes the following form,
\begin{equation}
\langle H_\alpha(t)H_\alpha(0)\rangle=\frac{\Delta^2}{\gamma_\mu^2}\exp(-\nu t).
\end{equation}
\begin{figure}[t]
	\centering
  \includegraphics[width=0.95\linewidth]{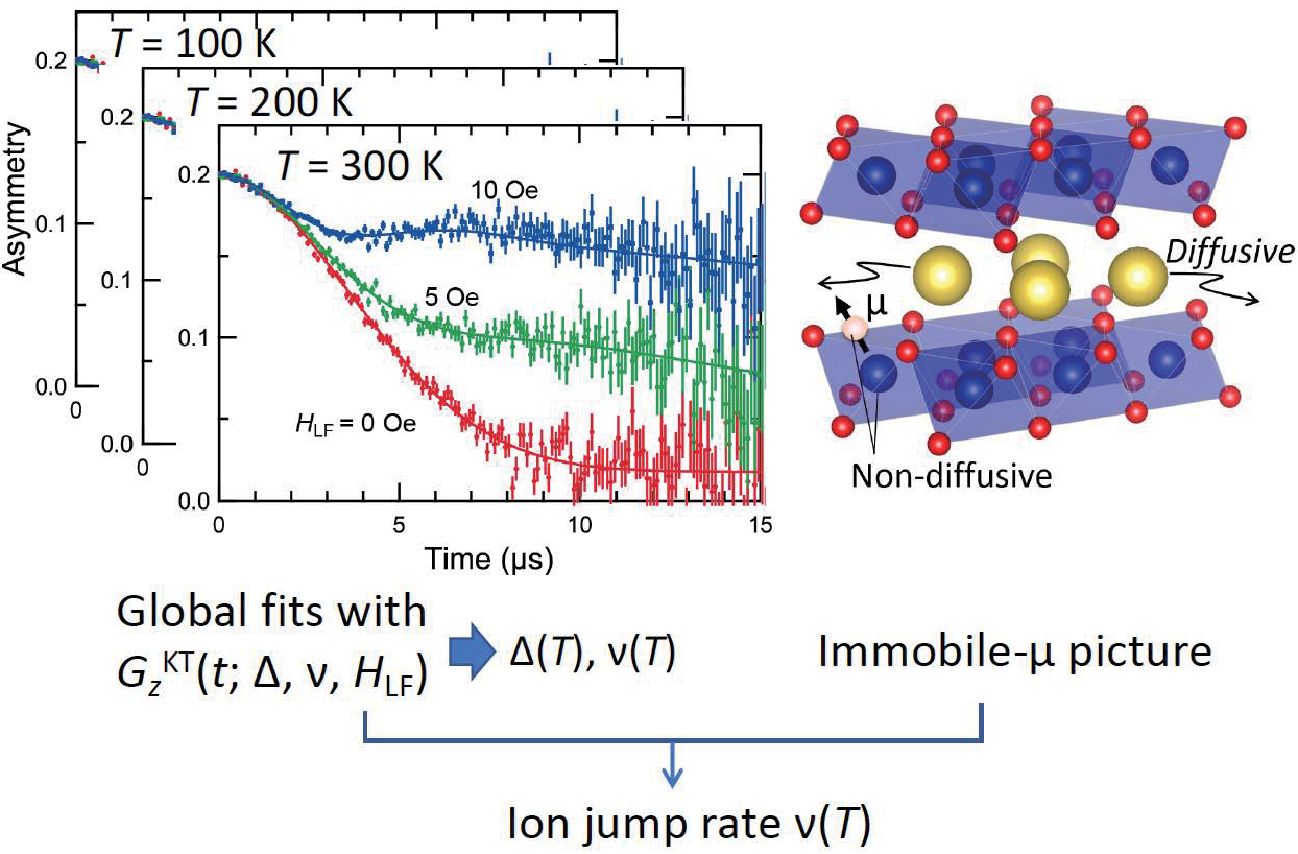}
\caption{Conceptual diagram of the conventional procedure for ion dynamics analysis using dKT function. Left: A set of $\mu$SR spectra at $H_{\rm LF}=0$, 5, and 10 Oe are analyzed by global fit to deduce $\Delta$ and $\nu$ at a given temperature $T$. (NB: The time spectra are simulations assuming the dKT function.) Right: Crystal structure of NaCoO$_2$, where muon ($\mu$) is located near oxygen (red spheres) and probes magnetic dipolar fields from $^{23}$Na (yellow spheres) and $^{59}$Co nuclei (blue spheres). }
\label{gfit}
\end{figure}
The dynamical relaxation function derived by Kubo and Toyabe was later modified into a form more suitable for describing abrupt change in ${\bm H}(t)$ associated with the muon self-diffusion [Fig.~\ref{fig2}(a)] \cite{Hayano:79}. In this formulation, the magnetic field fluctuations are treated as uncorrelated changes in the magnetic field vector that occur suddenly with frequency $\nu$ (strong collision approximation), and the corresponding dynamical relaxation function $G_z^{\rm KT}(t;\Delta,\nu)$ is obtained by numerically solving the following integral equation
\begin{eqnarray}
G_z^{\rm KT}(t;\Delta,\nu)&=&e^{-\nu t}g_z^{\rm KT}(t;\Delta)\nonumber\\ 
&+&\nu\int_0^t d\tau e^{-\nu(t-\tau)}g_z^{\rm KT}(t-\tau;\Delta)G_z^{\rm KT}(t;\Delta,\nu).\label{dKT}
\end{eqnarray}
Figure \ref{fig4} shows $G_z^{\rm KT}(t;\Delta,\nu)$ for several $\Delta/\nu$, which in the static limit $\Delta/\nu\rightarrow\infty$) coincides with the static KT function $g_z^{\rm KT}(t;\Delta)$ characterized by a ``1/3 tail''; the response with increasing $\nu$ first appears as a decay in the 1/3 tail (i.e., the increase in the relaxation rate of this component). As $\nu$ increases further, the overall shape of the curve changes from Gaussian to exponential with a decrease in the relaxation rate, approaching $G_z^{\rm KT}(t;\Delta,\nu)=1$ in the limit of $\Delta/\nu\rightarrow0$ (motional narrowing). Note here that the curve can vary over a wide range when $\nu$ changes from zero to infinity. As we will show later, in the new relaxation function we obtained by extending the dKT model, this range of variation is strongly suppressed \cite{Ito:24}. 

It is known that the dKT function formulated under the strong collision approximation behave almost identically with that under the Gaussian-Markov process except for slight differences in the slow fluctuation region ($\Delta/\nu\gg1$) \cite{Hayano:79}. Therefore, when we simply refer to the dKT function, we are mostly referring to those derived by the former that is widely used in the analysis of ion dynamics in solids as well. In any case, the point here is that the dKT function is used for the ion dynamics analysis under the assumption that it is not muons but some ions with nuclear magnetic moments that undergo diffusive motion, which seems divergent from what was originally envisioned.

\begin{figure}[t]
	\centering
  \includegraphics[width=0.65\linewidth]{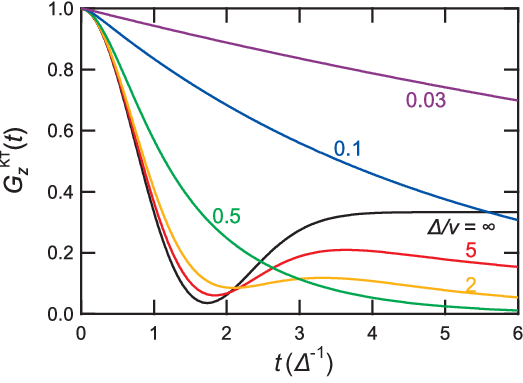}
\caption{Dynamical Kubo-Toyabe function $G_z^{\rm KT}(t;\Delta,\nu)$ at zero magnetic field, where curves are shown for several different $\Delta/\nu$.}
\label{fig4}
\end{figure}

While Eqs.~(\ref{sKT}) and (\ref{dKT}) are specific to spin relaxation in zero magnetic field, they can be readily extended to the cases in longitudinal magnetic field to obtain static relaxation function $g_z^{\rm KT}(t;\Delta,\nu,H_{\rm LF})$ and its dynamical version $G_z^{\rm KT}(t;\Delta,\nu,H_{\rm LF})$ \cite{Hayano:79}. In actual data acquisition and analysis, it is standard practice to determine $\Delta$ and $\nu$ as the common parameters in the simultaneous curve fits of several spectra obtained under different $H_{\rm LF}$ (as is found in Ref.\cite{Mansson:13}), which is called ``global fits'' [Fig.~\ref{gfit}].  This is because the $H_{\rm LF}$ dependence of $G_z^{\rm KT}(t;\Delta,\nu,H_{\rm LF})$ varies significantly with $\Delta/\nu$, allowing these two parameters to be reliably determined.
\begin{figure*}[t]
	\centering
  \includegraphics[width=0.72\linewidth]{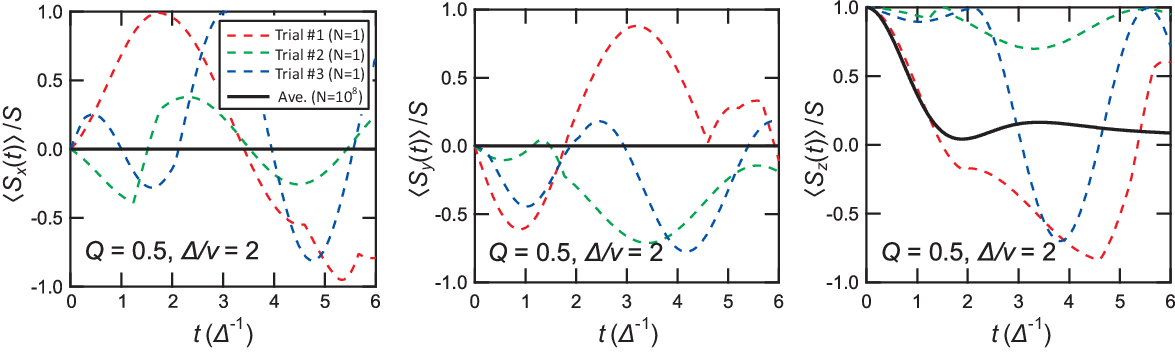}
\caption{Time evolution of the $x$, $y$, and $z$ components of $\langle{\bm S}(t)\rangle/|{\bm S}|$ under the local magnetic field specified by the parameters ($Q=0.5$, $\Delta/\nu=2$, $H_{\rm LF}=0$). Initial values were set to $(S_x,S_y,S_z)=(0,0,1)$ to match the $\mu$SR experiment arrangement under zero and a longitudinal magnetic field. The dashed lines represent the results of three independent trials, and the solid lines represent the average over $10^8$ trials.}
\label{fig5}
\end{figure*}

The parameters $\Delta$ and $\nu$ obtained in this way are further analyzed as a function of temperature and finally linked to parameters of ion diffusion. The linewidth $\Delta$ contains information about the position of the muon in the lattice, since it is a quantity determined by the magnetic dipolar fields that are proportional to $1/|{\bm r}|^3$ and thus sensitive to the distance $|{\bm r}|$ between the muon and the surrounding nuclear magnetic moments. The fluctuation rate $\nu$ is an important physical quantity that is interpreted as the jump frequency of the ions, and is tied to the attempt frequency of the ion jump $\nu_0$ in a potential well and the activation energy $E_{\rm a}$ through the Arrhenius relation, $\nu(T)=\nu_0\exp(-E_{\rm a}/k_BT)$. However, the reported values of $\nu_0$ in the earlier $\mu$SR literature are mostly on the order of $10^6$ to $10^7$ s$^{-1}$ \cite{Sugiyama:09,Mansson:13,Baker:11,Sugiyama:20,Ohishi:22a,Umegaki:22,Ohishi:23}, which is about six orders of magnitude smaller than those predicted by standard theories of classical ion diffusion \cite{Mehrer:07,Shewmon:16}. No reasonable explanation for this discrepancy has yet been provided. In addition, since the conventional analysis procedure assumes that the muon is immobile with respect to the lattice, the presence of stationary atoms on the lattice with a nuclear magnetic moment necessitates consideration on the contribution of the static local magnetic field [Fig.~\ref{fig2}(b)]. The use of the dKT function in this context means that this static contribution is ignored, questioning the credibility of the conventional procedure. In the next section, a new relaxation function that incorporates the effect of the static component will be derived using Monte Carlo calculations, and this will be used to verify this issue.

\section{Derivation of the relaxation function considering the contribution from ions fixed at lattice points}
Let us consider the case where static and dynamical sources of ${\bm H}$ coexist microscopically and each of them independently obey three-dimensional Gaussian distributions with standard deviation $\Delta_{\rm s}/\gamma_\mu$ and $\Delta_{\rm d}/\gamma_\mu$. Then ${\bm H}$ is given by a sum of two vectors, ${\bm H}_{\rm s}$ and ${\bm H}_{\rm d}$, randomly chosen from those representing static and dynamical magnetic fields. When dealing with longitudinal magnetic fields, we can further consider the field vector ${\bm H}_{\rm LF}$. Since the relaxation function for $\Delta_{\rm s}=0$ should coincide with $G_z^{\rm KT}(t;\Delta_{\rm d},\nu,H_{\rm LF})$, the fluctuations of ${\bm H}_{\rm d}$ should be treated as in the dKT model. Specifically, this can be incorporated by changing only ${\bm H}_{\rm d}$ at random with frequency $\nu$ and updating ${\bm H}$. In this case, ${\bm H}(t)={\bm H}_{\rm s}+{\bm H}_{\rm d}(t)$ exhibits non-vanishing autocorrelation as $t\rightarrow\infty$, which can be expressed approximately by that in the Edwards-Anderson theory for spin glass \cite{Edwards:76,Uemura:85},
\begin{eqnarray}
\langle H_\alpha(t)H_\alpha(0)\rangle&\approx&\frac{\Delta_{\rm s}^2}{\gamma_\mu^2}+\frac{\Delta_{\rm d}^2}{\gamma_\mu^2}\exp(-\nu t)\nonumber\\
 &=& (1-Q)\frac{\Delta^2}{\gamma_\mu^2}+Q\frac{\Delta^2}{\gamma_\mu^2}\exp(-\nu t),\label{EA}
\end{eqnarray}
where $\Delta^2$ ($=\Delta_{\rm s}^2+\Delta_{\rm d}^2$) is the total linewidth and $Q$ is the  parameter representing the relative weight of the dynamical component \cite{Ito:24} (see Fig.~\ref{fig2}(b), right). (Note that the meaning of the coefficients $1-Q$ and $Q$ in Eq.~(\ref{EA}) are opposite to that in the previous literature \cite{Edwards:76,Uemura:85}.)

Practically, it turns out that the integral equation approach as in Eq.~(\ref{dKT}) is not convenient for calculating the relaxation function with $0<Q<1$. Therefore, we adopted Monte Carlo simulations to obtain the desired function $G_z^{\rm EA}(t;Q,\Delta,\nu,H_{\rm LF})$ (denoted $G_z^{\rm ID}$ in Ref.~\cite{Ito:24}). The idea is to stochastically generate ${\bm H}(t)$ felt by an individual muon, including its time variation, and to track the precession of muon spin ${\bm S}(t)$ in three dimensions, which are finally averaged over a large number of muons. The dashed line in Fig.~\ref{fig5} depicts the time evolution of its expected value $\langle{\bm S}(t)\rangle/|{\bm S}|$ for one muon spin for each of three independent trials. The kink appearing intermittently on the sinusoidal curve corresponds to the moment when the direction and magnitude of ${\bm H}(t)$ is reset by the change in ${\bm H}_{\rm d}$. The average of this seemingly random set of curves converges to the smooth curve in Fig.~\ref{fig4} as the number of trials increases. We performed a similar calculation for each grid point in the four dimensional parameter space ($Q$, $\Delta$, $\nu$, and $H_{\rm LF}$) with $10^8$ trials and obtained a numerical table of relaxation functions $G_z^{\rm EA}(t;Q,\Delta,\nu,H_{\rm LF})$.

\begin{figure}[b]
	\centering
  \includegraphics[width=0.95\linewidth]{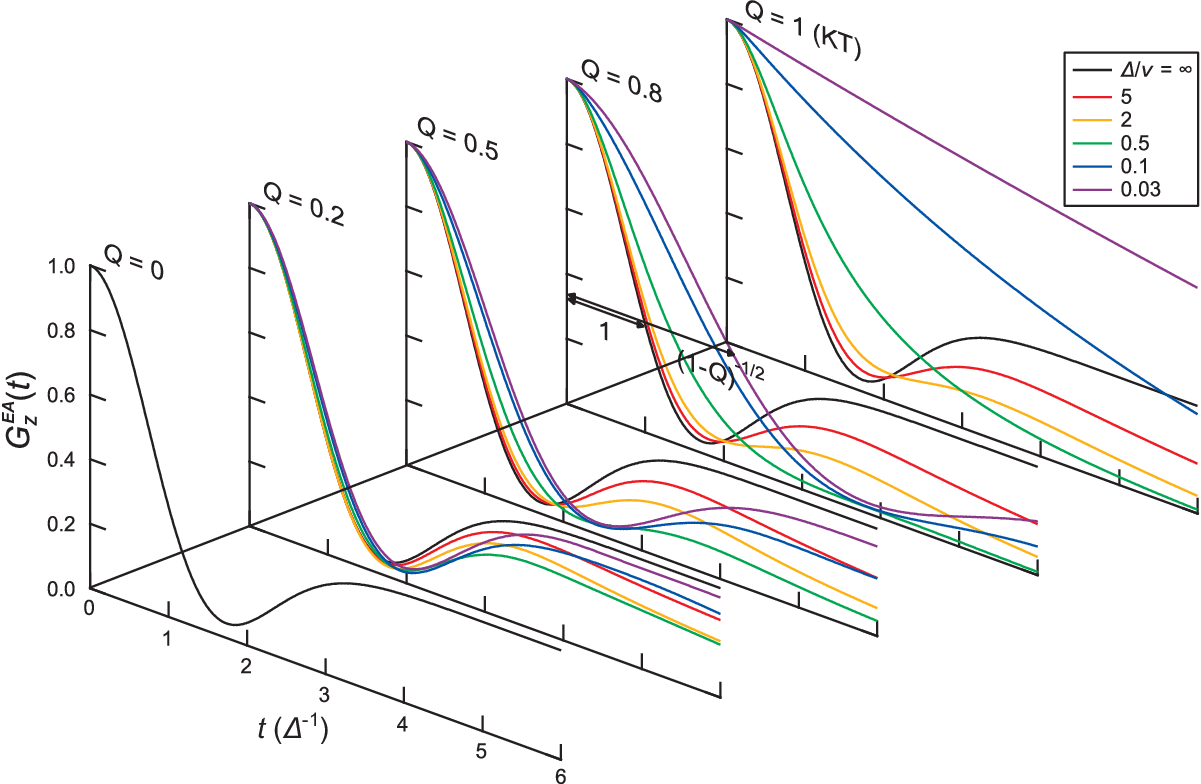}
\caption{Typical examples of the extended dKT function $G_z^{\rm EA}(t;Q,\Delta,\nu,H_{\rm LF})$ at $H_{\rm LF}=0$ with $Q=0$, 0.2, 0.5, 0.8 and 1, where curves are given for $\Delta/\nu=0.03$, 0.1, 0.5, 2, 5, and infinity at each $Q$. The decrease in the effective linewidth from $\Delta$ to $\sqrt{1-Q}\Delta$ with $\Delta/\nu\rightarrow0$ is clearly seen at $Q=0.8$. }
\label{fig6}
\end{figure}
Figure 6 shows the behavior of the extended dKT function $G_z^{\rm EA}(t;Q,\Delta,\nu,H_{\rm LF})$ under a zero magnetic field ($H_{\rm LF}=0$): for $Q=0$, only the static component remains ($\Delta_{\rm d}=\sqrt{Q}\Delta=0$), so that the relaxation curve is consistent with the static KT function given by Eq.~(\ref{sKT}). On the other hand, when $Q=1$, only the dynamic component remains ($\Delta_{\rm s}=\sqrt{1-Q}\Delta=0$), so the relaxation curve agrees with the dKT function based on the strong collision approximation in Eq.~(\ref{dKT}). Thus, it is confirmed that  $G_z^{\rm EA}(t;Q,\Delta,\nu)$ is reasonably connected to the existing model in the both ends of $Q$. Next, let us examine the behavior when $Q$ varies from zero to unity: in the region where $Q$ is small, the static component is dominant, and the corresponding static KT function-like behavior prevails over the entire time range. This trend does not change even if $\nu$ is varied from zero to infinity. The spectrum drifts only within a narrow range of amplitude, in contrast to the dKT function ($Q = 1$) that exhibits a large deviation from static KT function with varying $\nu$. When the linewidth is comparable between static and dynamical components ($Q = 0.5$), the range of deviation increases slightly and the 1/3 tail exhibits clear damping. However, the lineshape as a whole does not show any sign of motional narrowing for $\Delta/\nu\rightarrow0$. This tendency remains the same at $Q=0.8$, where it can be seen that the relaxation curve asymptotically approaches $g_z^{\rm KT}(t;\sqrt{1-Q}\Delta)$ when $\Delta/\nu\rightarrow0$. This means that even when the contribution of the dynamic component is averaged out and practically disappears due to fluctuations, the remaining static component (with a linewidth $\sqrt{1-Q}\Delta$) dominates the change in the lineshape at the motional narrowing limit, demonstrating the importance of considering the contribution of the static component in a zero magnetic field. This is quite reasonable, recalling that the zero field is the condition that provides the highest sensitivity to ``zero frequency noise'' \cite{Slichter:89}.

\section{Comparison with experiment}
Now that we are well-informed about the qualitative behavior of the extended dKT function, our task is to analyze the actual $\mu$SR data with this function and to compare the results with those based on the conventional analysis procedure. To this end, Li$_x$CoO$_2$ would comprise a typical example of ion dynamics in solids studied by $\mu^+$SR, for which significant amount of experimental knowledge is accumulated. However, since a coherent understanding is yet to be reached on the kinetic state of muons in this material (as there are mixed interpretations based on the assumptions in which muons are either immobile \cite{Sugiyama:09} or diffusing \cite{Ohishi:22b}), we do not deal with it here. 

Instead, we examine the $\mu$SR results on the cathode material of Na ion batteries, Na$_{0.7}$CoO$_2$, because the assumption of immobile muon has been consistently employed in this compound \cite{Mansson:13}. Moreover, additional studies using $\mu^-$SR is helpful to reduce uncertainty in the ion dynamics \cite{Sugiyama:20}; the implanted $\mu^-$ forms a spin-polarized muonic oxygen atom ($\mu^-$O) that provides information on the local magnetic field at the oxygen site [9]. Since the value of $Q$ is determined by the positional relationship between the probe and the ions,  $\mu^+$SR and $\mu^-$SR provide independent data sets with different $Q$ which are used to examine the mutual consistency in their interpretations.

\begin{figure}[t]
	\centering
  \includegraphics[width=0.97\linewidth]{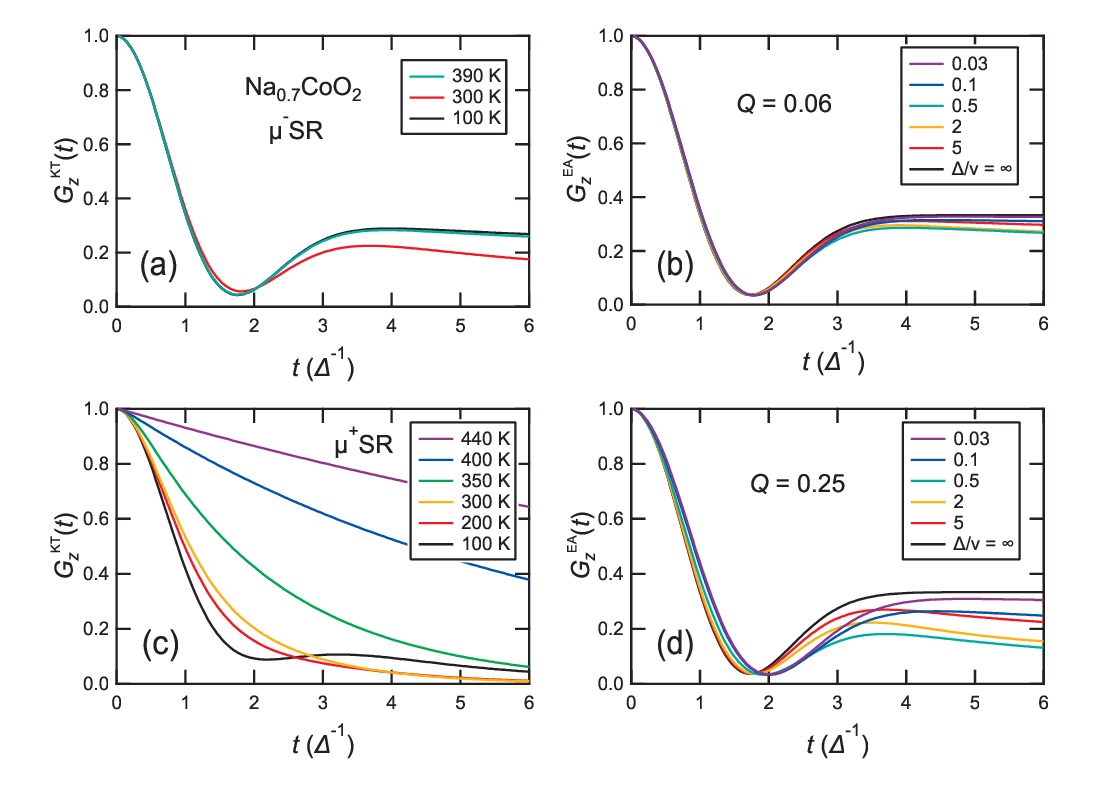}
\caption{(a) Best-fit curve obtained by analyzing the ZF-$\mu^-$SR spectra in Na$_{0.7}$CoO$_2$ with the dKT function \cite{Sugiyama:20}. (b) Lineshape of extended dKT function $G_z^{\rm EA}(t)$ at $Q = 0.06$. (c) Best-fit curve obtained by analyzing the ZF-$\mu^+$SR spectra with the dKT function \cite{Mansson:13}. (d) Lineshape of $G_z^{\rm EA}(t)$ at $Q = 0.25$.}
\label{fig7}
\end{figure}

First, let us examine the $\mu^-$SR part, where data have been obtained for 100--390 K and analyzed by  the conventional procedure shown in Sect.~2, to deduce the temperature dependence of $\Delta$ and $\nu$ \cite{Sugiyama:20}. The ZF-$\mu$SR spectra corresponding to the parameter values at 100, 300, and 390 K are reproduced in Fig.~\ref{fig7}(a). It is expected that the jump frequency of Na ions comes into the sensitive time range of $\mu$SR above $\sim$300 K, but these spectra exhibit least variation with temperature, confirming a static behavior characterized by a 1/3 tail over the entire temperature range. This seemingly unexpected behavior is interpreted to be due to the relatively small contribution from the Na ions to the local magnetic field. 
According to a detailed description on the estimation of $\Delta$ in the literature \cite{Sugiyama:20}, the contribution from the $^{59}$Co nuclei at the $\mu^-$O site is about 97\% of the entire $\Delta$.

Now, the above situation can be expressed by Eq.~(\ref{EA}), where the EA parameter can be obtained by the equation $\sqrt{1-Q} = 0.97$, which yields $Q = 0.06$. The corresponding lineshape of $G_z^{\rm EA}(t; Q=0.06,\Delta,\nu)$ ($H_{\rm LF}=0$) is shown in Fig.~\ref{fig7}(b), which hardly changes even when $\nu$ is varied over the entire range from zero to infinity, keeping a static KT function-like shape. Thus,  it is confirmed to be difficult to extract information on Na ion diffusion by $\mu^-$SR for Na$_{0.7}$CoO$_2$, as has been discussed previously  \cite{Sugiyama:20}. This also ensures that the $^{59}$Co nuclear magnetic moments are stationary on the $\mu$SR time scale, which will be important when discussing the $\mu^+$SR results later.

The results of $\mu^+$SR and $\mu^-$SR experiments on Na$_{0.7}$CoO$_2$ are summarized in the literature \cite{Mansson:13,Sugiyama:20}, where data acquisition, analysis, and interpretation are reported to be conducted according to the the conventional procedure. Here, typical ZF-$\mu$SR time spectra with $\Delta$ and $\nu$ given in Ref.~\cite{Mansson:13} for the dKT function parameters are shown in Fig.~\ref{fig7}(c) to reproduce the experimental data, which confirms the progressive change in the lineshape due to motional narrowing toward higher temperatures. This is in contrast to the $\mu^-$SR experiment, where almost no change in temperature was observed. The discrepancy was attributed to the  difference in the contribution from the $^{23}$Na nuclear magnetic moments to ${\bm H}(t)$ at the probe sites between the two methods \cite{Sugiyama:20}. However, according to the reported results of first principles calculations and dipole field calculations, the contribution of $^{59}$Co nuclei at the $\mu^+$ site is $\sim$87\% of the total $\Delta$, indicating that the quasistatic contribution from the $^{59}$Co nuclei remains dominant. The situation corresponds to $\sqrt{1-Q} = 0.87$ in the extended dKT model, which gives $Q = 0.25$.  As shown in Fig.~\ref{fig7}(d), while the variation of $G_z^{\rm EA}(t; Q=0.25,\Delta,\nu)$ with $\nu$ is slightly wider in amplitude than that in the $\mu^-$SR case [Fig.~\ref{fig7}(b)], it is still dominated by the static KT function-like shape. This is a qualitatively different behavior from the experimental data for $\mu^+$SR reproduced in Fig.~\ref{fig7}(c), and suggests that there is a major inconsistency between the assumption of immobile muons (as modeled in $G_z^{\rm EA}(t)$) and the analysis using the dKT functions.

So which of these is the problem? Here, we will stand on the immobile muon picture and investigate whether there exists a possibility for $G_z^{\rm EA}(t)$ to reproduce the behavior of the time spectra at high temperatures. The key parameter is the linewidth for the static component, $\Delta_{\rm s}=\sqrt{1-Q}\Delta$, whose contribution in $G_z^{\rm EA}(t)$ remains in the limit of fast fluctuations. $\Delta_{\rm s}$ can be estimated by dipole field calculations considering only the contribution of $^{59}$Co nuclei. The magnitude of $\Delta_{\rm s}$ at the arbitrary position in the crystal was calculated and found to have a minimum value of $\Delta_{\rm min}=0.15$ $\mu$s$^{-1}$ at the Na site. This means that no matter where the probe is placed in the crystal, static KT relaxation with $\Delta_{\rm s}\ge\Delta_{\rm min}$ will always remain in the high-temperature limit. This is contrary to the experimental fact shown in Fig.~\ref{fig7}(c). Therefore, at least at high temperatures above 350 K, we are forced to presume a situation where the $^{59}$Co nuclear contribution appears dynamical from the muon's point of view. In contrast, $\mu^-$SR results ensure that the $^{59}$Co nuclear magnetic moments themselves are static on the $\mu$SR time scale. Therefore, it is more reasonable to assume that the $^{59}$Co contribution appears dynamical because of the jump motion of muons relative to the $^{59}$Co nuclei. This result suggests that the assumption of immobile muon, which has been widely adopted in the study of ion dynamics, is not evident in itself and that the physical picture of the entire system, including the kinetic state of the probe, must be carefully examined based on data from each experiment. The extended dKT function will not only support quantitative analysis on the muon stationary picture, but will also play an important role in examining the validity of this physical picture (i.e., distinguishing the origins of fluctuations in ${\bm H}(t)$).

As is also true for the example discussed here, when oxides are studied, the analogy between protons and $\mu^+$ is often mentioned as support for the validity of immobile muon picture. This is because $\mu^+$ implanted into an oxide is usually considered to be strongly attracted to oxygen to form OH-like bonded state. It should be added here, however, that in practice, there have been a few reported cases of muon diffusion becoming active at low temperatures due to quantum tunneling of $\mu^+$ and/or a significant lowering of the effective barrier reflecting pronounced zero-point oscillations \cite{Nishida:91,Hempelmann:98,Ito:17,Dehn:21,Ito:23}.

\section{Discussion and Summary}
Although this paper has focused on the {\sl qualitative} difference between the $G_z^{\rm EA}(t)$ function and the $\mu$SR data in Na$_{0.7}$CoO$_2$, we have developed a plug-in for the data analysis software musrfit \cite{musrfit} so that it can be used for the least-square curve fitting of experimental data. Readers are encouraged to obtain it from the relevant URL for use as appropriate \cite{edKT}. With the development of this plug-in, we believe that we have provided a means to enable quantitative analysis and evaluation of its validity in a manner consistent with conventional interpretations. 

However, a special attention should be payed to the general assumption employed for both dKT and  extended dKT models that the entire system including the muon is in thermal equilibrium, disregarding the possibility of time-dependent changes in the statistical properties of the system (e.g., the local field distribution $n({\bm H})$).  In reality, the time origin in $\mu$SR measurements is the special moment in that the sample undergoes local electronic excitations by the kinetic energy of muons. The system after muon implantation does not necessarily reach thermal equilibrium immediately. In fact, numerous examples have been observed for muons to remain in a metastable state for a long time at low temperatures in a variety of materials \cite{Kadono:24}. Such a state may be more likely to occur in ionic conductors, which contain randomness due to the distribution of ion vacancies. As the temperature is increased, the system eventually relaxes toward the thermal equilibrium distribution due to thermal fluctuations. For example, in $\mu^+$SR experiments on VO$_2$, a decrease in $\Delta$ with increasing temperature (in striking resemblance to that in Li$_x$CoO$_2$ \cite{Sugiyama:09}) has been observed, which has been interpreted as an increase in the average distance between muons and $^{51}$V nuclear magnetic moments due to the diffusion-limited trapping of muons into the ion vacancies/voids \cite{Okabe:24}. In the temperature regime where the time characterizing the relaxation of the muon site/state distribution overlaps with the $\mu$SR time scale, it would be necessary to consider more specific models that incorporate the time-dependent  changes probed by muons (such as the two-state model \cite{Hempelmann:98,Ito:17,Ito:23}).

Furthermore, although rarely discussed in the context of ion dynamics analysis in solids, both positive and negative muons embedded in a sample can create large electrostatic disturbances to the surrounding ions. Thus, the mobile ions and muons are expected to strongly influence each other, so that when one is stationary, it appears to be a repulsive or attractive center to the other, and when both are in diffusive motion, a strong correlation in their motion is expected to emerge. To understand this effect, it will be important to analyze it from a molecular dynamics perspective. Although recent advances in computers and software have made it easy for anyone to try molecular dynamics calculations, more sophisticated and costly methods such as path integral molecular dynamics \cite{Miyake:98,Kimizuka:18} are essential to accurately treat systems including light muons, since the quantum nature of the ``nucleus'' must be taken into account. 

Thus, there are still many issues that need to be considered before the application of $\mu$SR is established as a true quantitative analysis technique for ion dynamics in solids. We hope that the model presented in this paper will provide a starting point toward solving these problems.

\section*{Acknowledgment}
We would like to thank K. Fukutani for his valuable advice and support. The Monte Carlo simulation was conducted using the supercomputer HPE SGI8600 in Japan Atomic Energy Agency (JAEA).
This work was financially supported in part by JSPS Grants-in-Aid for Scientific Research (Grant No.~24H00477, 23K11707, 21H05102, and 20H01864), and by the MEXT Program: Data Creation and Utilization Type Material Research and Development Project (Grant No. JPMXP1122683430). This paper is based on a topical article  published in the Japanese journal ``Kotai Butusuri (Solid State Physics)'' \cite{Ito:25}. 
\let\doi\relax

\begin{thebibliography}{32}
\bibitem [{Com()}]{Comt}%
  \BibitemOpen
  \href@noop {} {}\bibinfo {note} {In a Gaussian-Markov process, $n({\bm H})$
  itself is governed by random forces, and there assumed to be a correlation
  between ${\bm H}(t_i)$ and ${\bm H}(t_{i+1})$ according to $n({\bm H})$.
  (This can be understood by replacing ${\bm H}$ with the position ${\bm r}$ of
  a particle that undergoes Brownian motion.) On the other hand, in the strong
  collision approximation, $n({\bm H})$ is assumed to be a distribution
  independent of the random force, and ${\bm H}(t_{i+1})$ is allowed to take
  any value that follows $n({\bm H})$, independent of ${\bm
  H}(t_{i})$.}\BibitemShut {Stop}%
\end{thebibliography}
%

\end{document}